\title{Primordial RNA Replication\\and Applications in PCR Technology}
\author{
Stan Palasek\\
Princeton University
}
\date{May 2013}

\documentclass[12pt]{article}
\usepackage[margin=1in]{geometry}
\usepackage{setspace}
\usepackage{amsfonts}
\usepackage{amssymb}
\usepackage{amsmath}
\usepackage{wasysym}
\usepackage{graphicx}
\usepackage{wrapfig}
\begin{document}

\maketitle

\begin{abstract}
The emergence of self-replication and information transmission in life's origin remains unexplained despite extensive research on the topic. A hypothesis explaining the transition from a simple organic world to a complex RNA world is offered here based on physical factors in hydrothermal vent systems. An interdisciplinary approach is taken using techniques from thermodynamics, fluid dynamics, oceanography, statistical mechanics, and stochastic processes to examine nucleic acid dynamics and kinetics in a hydrothermal vent from first principles. Analyses are carried out using both analytic and computational methods and confirm the plausibility of a reaction involving the PCR-like assembly of ribonucleotides. The proposal is put into perspective with established theories on the origin of life and more generally the onset of order and information transmission in prebiotic systems. A biomimicry application of this hypothetical process to PCR technology is suggested and its viability is evaluated in a rigorous logical analysis. Optimal temperature curves begin to be established using Monte Carlo simulation, variational calculus, and Fourier analysis. The converse argument is also made but qualitatively, asserting that the success of such a modification to PCR would in turn reconfirm the biological theory.
\end{abstract}

\section{Introduction}
The mutual dependence of proteins and nucleic acids in their replication leads to a catch-22 in the origin of life: the first organism would require enzymes to replicate its genome, but these enzymes cannot be synthesized without the nucleic acids that code for them.\ \cite[p.~223-4]{weaver} This paradox was first resolved with Thomas Cech's discovery of self-splicing RNA in the protozoan \textit{Tetrahymena thermophila}.\ \cite{kruger} The existence of such a ``ribozyme'' that can both carry genetic information and perform catalysis removes the necessity of an event during which proteins and nucleic acids spontaneously emerged. Thus, the idea that an RNA world preceded the modern DNA-RNA world, once speculation by Crick and his colleagues, \cite{crick,orgel} had reached the mainstream as the ``RNA world hypothesis.'' \cite{gilbert}

According to Cech, an RNA world transitioned to a ribonucleoprotein (RNP) world of protein-bound nucleotides which were favorable for their greater versatility in catalysis. A final transition led to the current DNA-RNA world along with the last universal common ancestor (LUCA) to take advantage of DNA's greater resistance to hydrolysis.\ \cite{cech} The remaining problem is the origin of the ribozymes themselves. By the early 1990s, leading molecular biologists rejected ``the myth of a self-replicating RNA molecule that arose \textit{de novo} from a soup of random polynucleotides.'' \cite{joyceorgel} Biophysicists of the last decade seem to have ignored this submission and went on to evaluate the plausibility of spontaneous assembly in the contexts of thermodynamics, statistical mechanics, and information theory. (e.g. \cite{baaske,chen7,obermayer,chen12,krammer}) This paper will proceed in the spirit of the cited physical approaches, particularly Obermayer 2011 where the objective is to use mathematics and computer simulation to identify qualitatively new phenomena. The premise here, as it was there, is that information stored in an RNA sequence is lost upon hydrolytic cleavage so somehow a lasting memory must have emerged, likely in the form of a replicator. It is imprudent to be overly concerned with the numerical details due to our ignorance of the chemical and thermal factors of ancient oceans.\ \cite{pagani, pinti} Also in the manner of the cited literature, the origin of the nucleotide monomers themselves will not be addressed as this is a different problem entirely (see for instance \cite{powner} and the assumptions made in \cite{obermayer,robertson}).

\section{Nucleotide Flow in a Hydrothermal Vent}

\subsection{Hydrothermal gradients}

Let the local energy per unit distance at a height $z$ above a hydrothermal vent  be $U(z)$. The generic scalar transport equation in steady state becomes Poisson's equation \cite[p.~14]{shubin}, $\alpha\nabla^2U+P(\textbf{r})=0$ where $\alpha$ is the thermal diffusivity and $P$ is the instantaneous power per unit distance. The hydrothermal vent is treated as a point source with $P_{vent}=2P_V\delta(z)$ where $\delta(z)$ is the Dirac delta function evaluated at a height $z$ above the vent, requiring the factor of two so that the differential power integrates to $P_V$ on $\mathbb{R}^+$. With just the constant influx of hydrothermal energy, the system would never reach equilibrium. Let $U_0$ be the total thermal energy in steady state. Suppose that there is dissipation per unit distance proportional to the local energy density, so $P_{lost}=UP_V/U_0$ in such a way that the net power is 0. Poisson's equation in the vertical direction takes the form of the screened Poisson equation \cite[pp.~312-313]{fetter} with $\lambda=\sqrt{P_V/\alpha U_0}$ and $f(z)=(2P_V/\alpha)\delta(z)$. 
\begin{equation}\label{heat}
\alpha\frac{d^2U}{dz^2}+2P_V\delta(z)-P_V\frac{U}{U_0}=0
\end{equation}
A solution is desired for which $\int_0^{\infty}U(z)dz$ both converges and equals $U_0$. These two constraints, though not independent, uniquely solve the initial value problem, giving an exponential function with decay constant $\lambda$. Exploiting the linearity between temperature and and average thermal energy and defining the ambient ($z=\infty$) and vent ($z=0$) temperatures as $T_0$ and $T_V$ respectively yields the following temperature distribution.

\begin{equation}\label{expTemp}
T(z)=T_0+(T_V-T_0)\exp\left(-\sqrt{\frac{P_V}{U_0\alpha}}z\right)
\end{equation}

\subsection{A new model for thermophoresis}

Thermophoresis is the tendency of particles to move against temperature gradients. Its mechanism is up for debate and has been modeled primarily as an entropic phenomenon across only small gradients.\ \cite{duhr} Here a model is derived based on the principle that more collisions with a test particle occur from the direction of higher temperature. Let the test particle occupy a van der Waals region $\mathcal{R}$ with a surface $\partial\mathcal{R}$ and volume $V$. By the ideal gas law, the ratio of the pressure orthogonal to the surface to the temperature is $k_BC$, where $k_B$ is Boltzmann's constant and $C$ is the local concentration of the solvent. Integration over the surface provides the total resulting force.
\begin{equation}\label{surfaceInt}
\textbf{F}\hspace{-.02in}_{thermal}=-\oiint_{\partial\mathcal{R}}k_BCT(x,y,z)\,d\hspace{-.02in}\textbf{A}
\end{equation}
It is shown\footnote{Assume $\mathcal{R}$ is compact and has a piecewise smooth boundary. Applying the divergence theorem to a vector field $\textbf{F}(x,y,z)=T(x,y,z)\,\textbf{c}$ where $\textbf{c}$ is a constant vector eventually gives $\textbf{c}\cdot\left(\oiint_{\partial\mathcal{R}}Td\textbf{A}-\iiint_R\nabla TdV\right)=0$ for all $\textbf{c}$.} that this surface integral can be converted to an integral over the volume. The limit is taken as the molecule's size approaches zero and the temperature gradient becomes approximately uniform throughout the region.
\begin{equation}\label{netF}
\eqref{surfaceInt}=-k_BC\lim_{\|\mathcal{R}\|\rightarrow 0}\iiint_{\mathcal{R}}\nabla TdV=-k_BCV\nabla T
\end{equation}
The force field due to thermophoresis is therefore conservative. This is equivalent to Archimedes' law with pressure replaced by the most probable (mode) ratio of thermal energy to volume. The constant is perhaps related to the Soret coefficient \cite[pp.~522-3]{cussler} of classical thermophoresis which has dimensions of inverse temperature rather than distance per temperature.

\subsection{Temperature oscillation}

The oscillations of interest are in the vertical direction over the hydrothermal vent, so consider \eqref{netF} in a single dimension superposed with constant acceleration of gravity $g$: $md^2z/dt^2=-k_BCV\partial_zT-mg$. Multiplying through by $dz/dt$ and integrating with respect to $dt$ explicitly gives the kinetic energy and gravitational potential energy as a function of temperature. Adding the work done by thermophoresis (integrating \eqref{netF}) yields the constant Hamiltonian $H=k_BCV(T_V-T_0)$. When oscillations are small and the right hand side of the trajectory equation is approximated with a first-order Taylor polynomial, each nucleotide behaves as a harmonic oscillator with angular frequency $\omega$.

\begin{equation}\label{traj}
x(t)=\left(\sqrt{\frac{U_0\alpha}{P_V}}-\frac{g}{\omega^2}\right)\left(1-\cos\omega t\right)\hspace{.4in}\omega^2=\frac{k_BCVP_V}{\alpha mU_0}(T_V-T_0).
\end{equation}

Figure 1 shows temperature as a function of time as a result of of these oscillations without the harmonic approximation. The shaded blue curve is the theoretical deterministic trajectory from the nonlinear equation obtained numerically with a Runge-Kutta method \cite[pp.~413-4]{hamming}. The scattered data depict four runs of the process taking into account not only thermophoresis and gravity, but also buoyancy, diffusion, and the temperature dependence of the transport coefficients \cite[p.~114]{cussler}. The existence of thermal cycling is critical to the rest of this paper.

\begin{figure}
\caption{Temperature oscillations from the perspective of a particle in a hydrothermal vent according to \eqref{netF} both deterministically (shaded blue) and stochastically (scattered).}
\hspace{-.05in}\includegraphics[scale=.765]{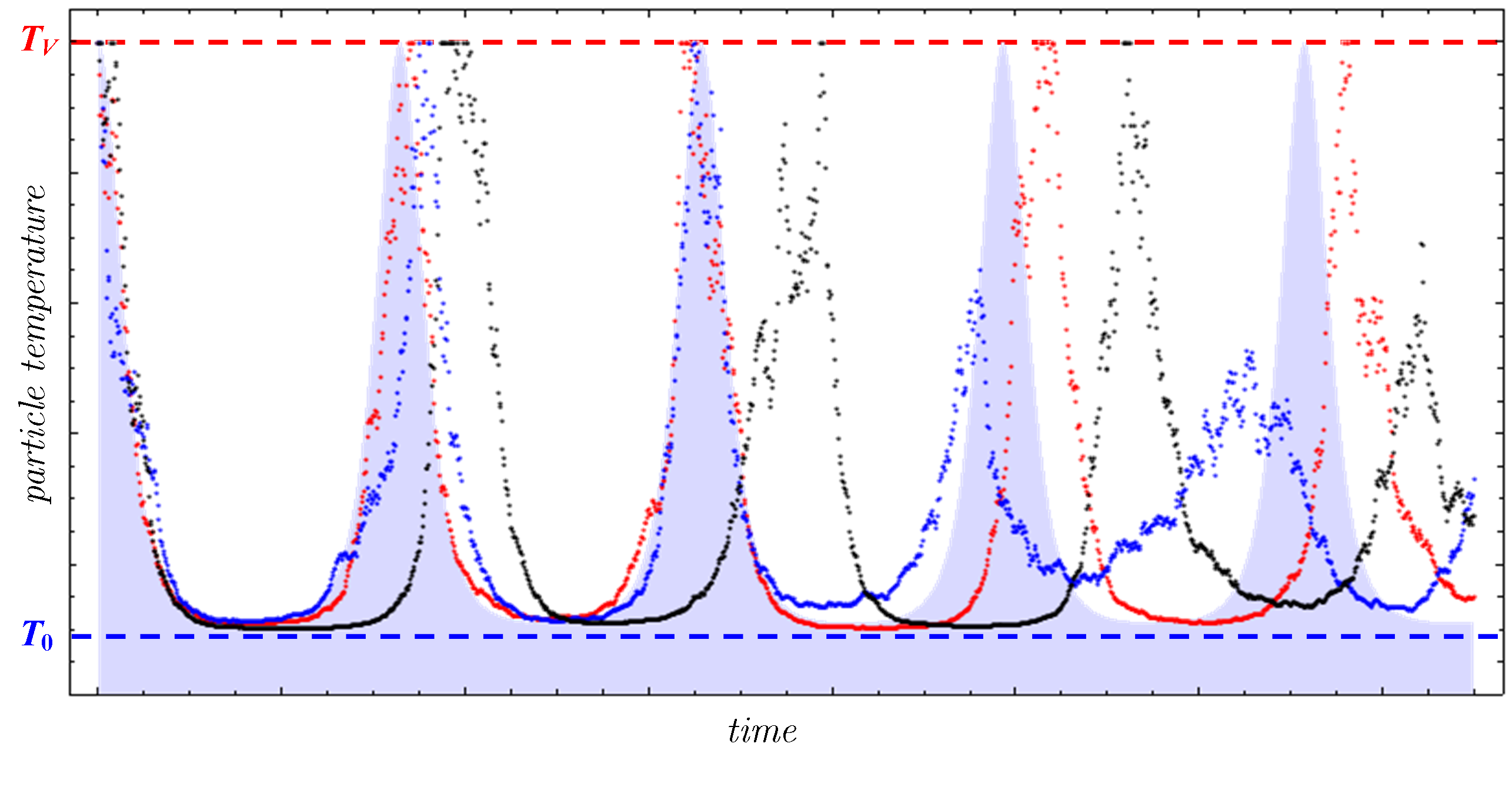}
\end{figure}

\section{The Quasi-Polymerase Chain Reaction Hypothesis}

In the previous section it was shown that particles in a hydrothermal vent system experience temperature oscillations. Here it is proposed that these oscillations could mirror the thermal cycling that is the fundamental basis of the polymerase chain reaction (PCR). A ``quasi-polymerase chain reaction'' (quasi-PCR) would consist of two primary stages: denaturation and rehybridization. The former stage takes place most favorably near the vent so that the hydrogen bonds united the complementary RNA strands may be destroyed. The latter stage must take place far from the vent at the ambient ocean temperature so that loci that have undergone rehybridization---a slow process when uncatalyzed---are not prematurely broken by rogue thermal fluctuations. It is essential to provide long spans of time at this low temperature. The oscillations shown in Figure 1 are merely to illustrate their existence and are too wide and too frequent to result from a hydrothermal vent that creates a temperature gradient potentially larger than 400$^\circ$C.\ \cite{haase} Hybridized molecules will actually stay at cooler temperatures longer than this model predicts by virtue of their size: the diffusion coefficient is inversely proportional to the particle's mass (also by the Stokes-Einstein equation \cite[\textit{eodem loc.}]{cussler}), fluid density and with it buoyancy increase as the water temperature decreases, and acceleration due to gravity is intensive. This proposal will be examined in more detail with mathematics and in computer simulation in subsection 5.3.

It should be emphasized that the author is \textit{not} suggesting that a bona fide PCR could progress in a hydrothermal vent. Indeed, it would be contrary to the motivating RNA world hypothesis to assume that polymerase (the \textit{P} in \textit{P}CR) enzymes would have yet been in existence in the same vent system. Without enzymes, the primers that delineate the nucleic acid fragment to be replicated are not needed and neither is an annealing step. It is acknowledged that the reaction's specificity and speed are compromised by removing these elements but it will be argued that only a simplified process is needed to lead in to an RNA world. The plausibility of the reaction without biological catalysts will be addressed in section 6.

\section{Analytic Modeling}

A description of an analytic model for the process described in the previous section will be presented before the Monte Carlo simulation as its consequences will be needed throughout the rest of the paper.

\subsection{Quasi-PCR as a branching process}

A method for examining the rate of polynucleotide branching in continuous time with discrete molecules will be developed. Let $f_n(t)$ be the probability mass function for the number of molecules ($n$) as a function of time ($t$). Branching of a given molecule occurs at a time-dependent rate $R(t)$, meaning the probability it branches (replicates) on an interval $[t,t+\Delta t]$ as $\Delta t\rightarrow0$ is $R(t)\Delta t$. Using a finite difference $\Delta t$, $f_n(t)$ can be expressed recursively as
\begin{equation}\label{rec}
f_n(t+\Delta t)=(n-1)f_{n-1}(t)R(t)\Delta t+f_n(t)\left[1-n R(t)\Delta t\right]\textnormal{ for }n=1,2,3,\ldots
\end{equation}
meaning the probability of being at $n$ on a given time step is the probability of being at $n-1$ on the previous time step and making a successful transition plus the probability of being at $n$ on the current time step and not making a successful transition. The probability a transition occurs is proportional to the number of loci involved.\footnote{The probability is actually given by $1-(1-r\Delta t)^n$ which, when $\Delta t\approx0$, has a Taylor approximation of $nr\Delta t$.} In the continuum limit, $\left[f(t+\Delta t)-f(t)\right]/\Delta t$ becomes $d\!f_n(t)/dt$.
\begin{equation}\label{diff}
\frac{d\!f_n(t)}{dt}=R(t)\left[(n-1)f_{n-1}(t)-nf_{n}(t)\right]=-R(t)[n\Delta(f_n(t))+f_{n-1}(t)]\textnormal{ for }n=1,2,3,\ldots
\end{equation}
adopting the notation of difference equations in the second representation. This system of countably infinite differential equations or, equivalently, a combined differential equation and recursive relation, has initial conditions $f_{0}(t)=0$ for all $t$ meaning that a system from non-zero intial conditions can never have zero particles and $f_n(0)=\delta_{n-1}$ for all $n$ where $\delta$ is the Kronecker delta meaning at $t=0$ the system is certain to contain only one molecule.\footnote{Other initial conditions can be accommodated simply by scaling the final probability mass function.} The general solution of the first-order linear equation is known, \cite{santos} allowing the representation of $f_n$ as an explicit function $f_{n-1}$.
\begin{equation}\label{branchonce}
f_n(t_n)=(n-1)\int_0^{t_n}\exp\left(-n\int_{t_{n-1}}^{t_n}R(k)dk\right)f_{n-1}(t_{n-1})R(t_{n-1})dt_{n-1}
\end{equation}
During process of the calculation, $t_n$ will be used as the argument of the function $f_n$ to disambiguate between the infinite independent variables.\footnote{All but $t_n$ turn out to be dummy variables anyway.} Each $f_k$ can be replaced iteratively by an expression in terms of $f_{k-1}$ using \eqref{branchonce} until $f_1$ is reached. Recalling that $f_0(t)=0$ and $f_1(0)=1$, \eqref{diff} yields $f_1(t)=\exp\left(-\int_0^tR(k)dk\right)$.
\begin{equation}
(n-1)!\int_0^{t_n}\cdots\int_0^{t_2}\exp\left(-\sum_{i=1}^{n-1}\int_{t_i}^{t_{i+1}}(i+1)R(k)dk\right)\exp\left(-n\int_0^{t_1}R(k)dk\right)\prod_{i=1}^{n-1}R(t_i)dt_1\cdots dt_{n-1}
\end{equation}
The integrals within the exponential do not simply combine because each one has a different coefficient. They can instead be combined into $n-1$ integrals with the same coefficients by combining the first through the $(n-1)$st, including the separate contribution from $f_1$, into a single integral over $[0,t_n]$; then the second to the $(n-1)$th into one on $[t_1,t_n]$; etc. Now the integrand can be concisely written as $\prod_{i=1}^{n-1}\exp\left(-\int_{t_i}^{t_n}R(k)dk\right)R(t_i)dt_i$ and the substitution $u_i=-\int_{t_i}^{t_n}R(k)dk$ can be made leaving only $\exp(u_1+\cdots+u_{n-1})du_1\cdots du_2$. From here it is found and can be easily verified by induction or by substitution into \eqref{rec} that $f_n(t_n)=\exp\left(-\int_0^{t_n}R(k)dk\right)\left[1-\exp\left(-\int_0^{t_n}R(k)dk\right)\right]^{n-1}$. Finally, we revert to $t$ without the subscript and rescale such that $f_n(0)=\delta_{n-n_0}$ where $n_0$ is the initial quantity of molecules.
\begin{equation}\label{branch}
f_n(t)=\exp\left(-n_0\int_0^tR(k)dk\right)\left[1-\exp\left(-n_0\int_0^tR(k)dk\right)\right]^{n-n_0}
\end{equation}
Recalling that $f$ is a probability mass function over the random variable $n$, if there were not a logical flaw, \eqref{branch} would be properly normalized.\footnote{A normalized result would be encouraging as nowhere in the derivation was it explicitly forced.} The distribution is further confirmed in Figure 2 in which the narrowness of the confidence interval (red) is a consequence of the very little deviation from the mean scaled by the square root of the sample size. Taking the sum from $n=n_0$ \textit{ad infinitum}, the exponential component is independent of $n$ and factors out of the sum, while all that remains is the sum of $(1-x)^n$ as $n$ goes from 0 to infinity. This is $1/x$ by the formula for a geometric series and happens to be the precise inverse of the exponential in \eqref{branch}. Thus it is normalized. To compute the expectation of the random variable $n$ as a function of time, compute the sum of $nf_n(t)$. Some manipulation gives
\begin{equation}\label{mean}
\langle n\rangle=n_0-1+\exp\left(n_0\int_0^tR(k)dk\right)
\end{equation}
where the angular brackets denote the expectation operator. It can be seen that \eqref{branch} is identical to the probability mass function for the shifted geometric distribution \cite[p.~68]{turin} shifted by $n_0$ rather than the conventional 1 and with $\lambda=\exp\left(-\int_0^tR(k)dk\right)$. This equivalence is only formal as nowhere in the formulation is there a notion of a Bernoulli process.

\begin{figure}[b!]
\caption{The dark line on the left is the number of molecules as a function of time predicted by \eqref{mean} when $R(t)=1+\sin(3t)$, surrounded by the 95\% confidence interval for the mean based on a sample of $10^4$ of such processes with $\Delta t=0.01$. The predicted and observed correspond closely. $R(t)$ is depicted below on the same time scale. To the right is the distribution of the number of molecules at the final time apposed with the geometric distribution given by \eqref{branch}.}
\hspace{-.05in}\includegraphics[scale=.55]{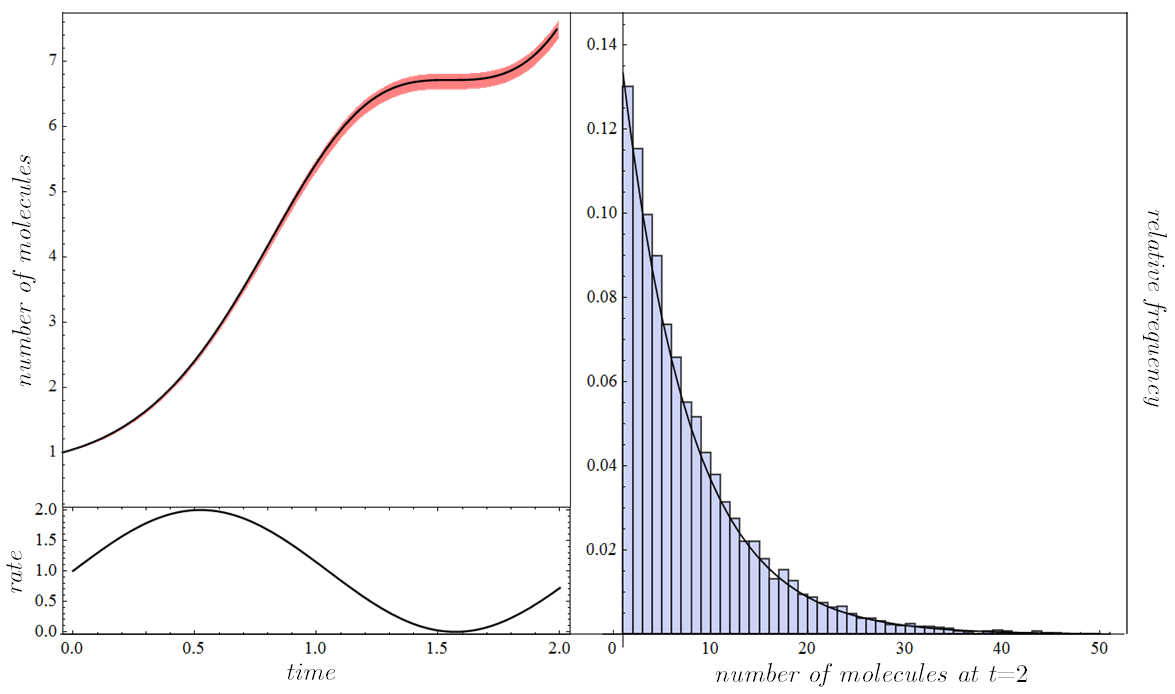}
\end{figure}

\subsection{Computing $R(t)$}

$R(t)$, the rate at which complete replications occur for a polynucleotide of length $m$, is (the probability that $m-1$ loci are hybridized at time $t$) $\times$ (the rate of hybridization of the final $m$th locus). The former probability is given simply by the binomial distribution, $m(1-p_0)p_0^{m-1}$ where $p_0$ is the probability that a given locus is hybridized. Under several circumstances this can be approximated by $mp_0^m$, including when $p_0\approx1/2\Rightarrow p_0\approx1-p_0$, when $l$ is large and edge effects are negligible, and when partial degradation of a fully replicated strand is infrequent. The latter justification can be put quantitatively as $p=(\textnormal{probability completely replicated})=p_0^m$ (assuming independence) such that when the reverse reaction is improbable $d\!p/dt=mp_0^{m-1}d\!p_0/dt$ and $d\!p_0/dt$ behaves as the rate of hybridization of the $m$th locus. One could then easily take into account the reverse reaction with $R-m\times(\textnormal{denaturation rate})=d\!p/dt$.

$p_0$ has two multiplicative components that are assumed to be independent: the probability the original double strand has been denatured and the probability it is in a hybridized state given it has been denatured. In order to make this problem tractable, we assume that the change in thermal energy is slow compared to the molecular and submolecular processes being examined here. The slowness that follows from this kind of adiabadicity would confine the system to a state infinitesimally close to equilibrium.\footnote{These assumptions are in line with those of the quasistatic field approximation that is likewise useful in electrical engineering.\ \cite[p.~31]{bansal}} This allows many of the processes considered here to be accurately treated as Poisson (memoryless).

First we compute the steady-state probability a locus is in a denatured state. This must be significantly lower for longer molecules since many loci will be quickly denatured and remain idle while the remainder of the molecule is being hybridized. Let $r_+$ and $r_-$ be the respective hybridization and denaturation rates, borrowing from the notation of \cite{obermayer}. The probability the Poisson event representing denaturation has occurred as of time $t$ from the beginning of a quasi-PCR cycle is $1-e^{-r_-t}$.\ \cite[p.~400]{turin} On average, the length of a cycle is the inverse of the turnover rate ($1/R$). The probability of a randomly chosen point in time being within the time interval $[t,t+dt]$ into the current cycle is $dt/(\textnormal{average cycle period})=Rdt$. Therefore, the expected probability of denaturation is strangely given by the integral of an already cumulative distribution, $\sum_{t\in\textnormal{period}}P[\textnormal{denatured~at~}t] \times P[t\textnormal{~into~the~cycle}]=\int_0^{1/R}\left(1-e^{-r_-t}\right)Rdt=1-\frac{R}{r_-}\left(1-e^{-r_-/R}\right)$. The memorylessness of hybridization-dehybridization that can be assumed in quasistatic equilibrium implies, finding the steady state of the simple two-state Markov chain, that the probability of being in a hybridized state is $1/(1+r_-/r_+)$. These results combined finally yield $p_0$ (again, assuming the independence of the probabilities).
\begin{equation}\label{p0}
p_0=\frac{1-\frac{R}{r_-}\left(1-e^{-r_-/R}\right)}{1+r_-/r_+}\sim\frac{1}{2R\left(r_+^{-1}+r_-^{-1}\right)}=\frac{r_{\textnormal{harmonic mean}}}{2R}\textnormal{  for non-negligible replication } (R\gg r_-).
\end{equation}
Solving this with $R-mr_-=mp_0^{m-1}d\!p_0/dt$ gives $R$ but is dependent on which rates are considered invariant with time. Specifics will not be given here in the context of quasi-PCR as these mathematics do not become useful until biotechnology applications are discussed toward the end of the paper where the kinetics fundamentally change.
 
\section{Monte Carlo Simulation of Quasi-PCR}

Due to the system's complexity, it is not practical to rely entirely on analytic methods. The implementation of a Monte Carlo simulation of the process outlined in Section 3 is detailed here.

\subsection{Implementing the hydrothermal gradient}

The temperature gradient is of magnitude $\frac{\partial T}{\partial z}=-\lambda(T_V-T_0)\exp\left(\lambda\,z\right)$ where $\lambda$ is again $\sqrt{\frac{P_V}{U_0\alpha}}$ and seems to act as a linear approximation for the exponential temperature gradient. The ratio $P_V/U_0$ is the ratio of the vent's power to the ocean's total potential energy relative to the ambient energy on the order of $k_BT_0$. It can thus be thought of as a turnover rate for energy in a vent system. $U_0$ can be estimated based on the temperature at the base of the vent. Direct solution of \eqref{heat} shows that the linear energy density at the base is $\sqrt{\frac{P_VU_0}{\alpha}}$. In a single dimension, the average kinetic energy of a single particle relative to the energy at the ambient temperature is $\frac{1}{2}\,k_B(T_V-T_0)$. Letting $\rho$ be the linear density of particles, we can equate the two energy densities, solve for $U_0$, and rewrite $\lambda=2P_V/[\alpha\rho k_B(T_V-T_0)]$. The remaining parameters must be empirical. Little \textit{et al.} examined hydrothermal vent flow in the East Pacific Rise, finding power $P_V=3.7\pm0.8$ MW.\ \cite{little} Thermal conductivity values for water at various high temperatures and pressures were established at the Sixth International Conference on the Properties of Steam.\ \cite{sengers} Vent temperatures and pressures on the Mid-Atlantic Ridge fall on or above seawater's critical point of 407$^\circ$ C and 29.8\,MPa.\ \cite{kos}

\subsection{Implementing thermophoresis}

To compute the particle dynamics as modeled by \eqref{netF}, the van der Waals volumes of nucleotides and water are computed using Zhao's method.\ \cite{zhao} Averaging over pyrimidine- and purine-derived bases and assuming Chargaff's rules apply \cite[p.~140]{weaver}, the constant factor in \eqref{netF} can be obtained with dimensions of specific (intensive) entropy. Rearranging the equation gives an acceleration due to thermophoresis $\textbf{a}=\left(421.5\frac{\textnormal{J}}{\textnormal{kg\,K}}\right)\nabla T$. In the limit of long polynucleotide length, molecular mass becomes proportional to van der Waals volume and the thermophoretic acceleration becomes independent of the length.

\subsection{Implementing diffusion}

The dynamics are modeled as a finite difference process with time step $\Delta t$ where the state of a particle at time $t$ is determined by its position $z_t$ and velocity $v_t$. On each step, the position is updated nondeterministically to $z_{t+\Delta t}=z_t+v_t\,\Delta t$. Then velocity is updated\footnote{Casual experimentation shows that when using a computationally-economical large time step, evaluating acceleration at $z_{t+\Delta t}$ rather than $z_t$ produces trajectories that diverge slower from those that are in the ideal continuum limit.} to $v_{t+\Delta t}=v_t+a(z_{t+\Delta t},l)\,\Delta t$ where $a$ is the acceleration that is explicitly dependent on both position and polymer length.  Particles diffuse in a Gaussian distribution centered about the average drift of the substance in bulk with variance $2Dt$ where $D$ is the diffusion coefficient.\ \cite[p.~37]{cussler} To achieve this, when computing $x_t$ at each step, a Gaussian random variable is added with variance $2D\Delta t$. Since the difference is finite and large thermal fluctuations can occur, it is necessary to reset the particle at $z=0$ if an anomaly puts it in an unphysical location.

The diffusion coefficient must vary dramatically with the temperature and molecule size. The standard Stokes-Einstein relation cited earlier will not be appropriate because it is restricted to spherical particles. Instead we will resort to the most general Einstein relation $D=\mu k_BT$ where $\mu$ is the particle's mobility.\ \cite{einstein} In the case of spherical particles and low Reynolds number the mobility would be replaced by the reciprocal drag coefficient which is given simply by Stoke's law as $\textit{drag}=6\pi\eta R$ where $\eta$ is the dynamic viscosity of the fluid and $R$ is the sphere's radius. However it would be na\"{i}ve to assume that $R$ scales with molecule size without going back to first principles. Ideally, one would begin with the Navier-Stokes equations and use a cylindrical coordinate system (where long molecules can be neatly represented) rather then a spherical one (which is most convenient computationally because the stream function can be decomposed into radial and polar components). This formulation is still in progress. Instead, consider the fact that precisely one-third of the drag force is due to pressure and the rest is the result of shear stress.\ \cite{mit} The magnitude of a pressure gradient is dependent on the vertical size of the molecule. This size can be modeled as the difference between the maximum and minimum heights reached in a random walk in continuous three-dimensional space with fixed step size (neglecting that a molecule is self-avoiding). A Monte Carlo simulation shows that this range has a least-squares fit of $l_Bl\,^{0.516}$ ($R^2>0.9999$, best fit in the asymptotic limit of large $l$ as shown in Figure 3) where $l_B$ is the length of a single segment. The viscosity, on the other hand, is a consequence of stress between the fluid and the molecule which must be proportional to the surface area. Nucleic acids with their negatively charged phosphate groups will conform to maximize this surface area, making it proportional to the molecule length. Weighting the dependencies on $l^{1/2}$ and $l$ based on their contributions to the total drag gives $\textit{drag}=2\pi\eta R(\hspace{-.02in}\sqrt{l}+2l)$ where $R$ is the radius of a single nucleotide approximated by a sphere. In the monomeric case, this formula correctly reduces to the Stokes-Einstein equation.

\begin{figure}
\caption{Average sizes of folded nucleotides modeled with a random walk, computing the size as a function of length up to length 1000 sampling 1000 conformations. The unit for distance is the length of a monomer. The log-log scale makes clear that the linear dimension asymptotically approaches precisely $\sqrt{m}$. This is confirmed in the residual plot in which the magnitude of the sampling error is correctly on the order of the square root of the observed value.}
\hspace{-.06in}\includegraphics[scale=.4095]{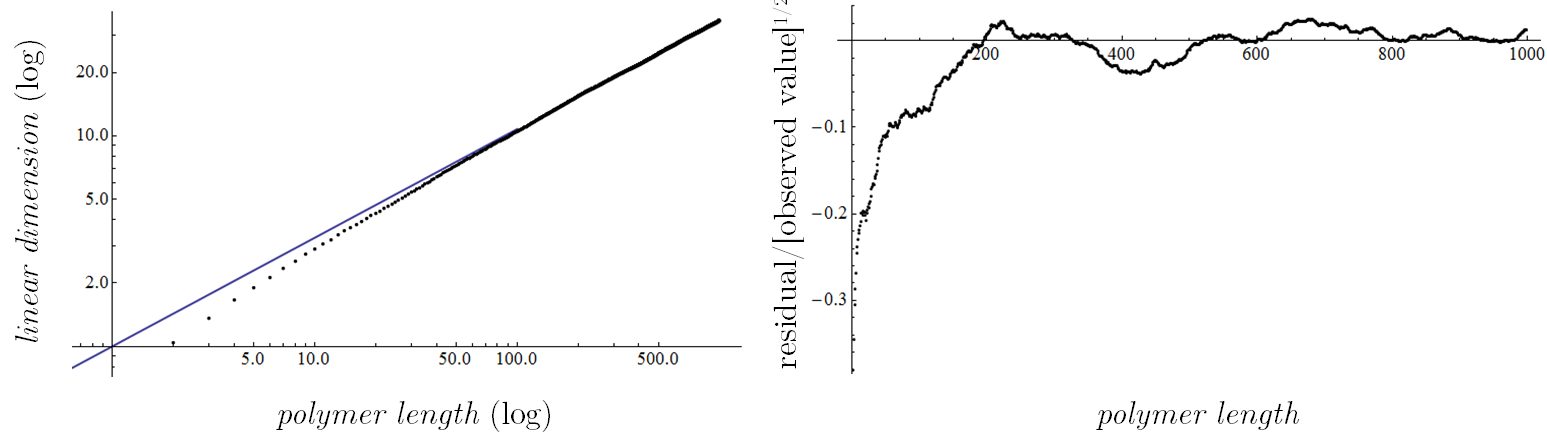}
\end{figure}

\subsection{Implementing bond kinetics}

Maxwell-Boltzmann statistics describe the probability that a given particle is in a state with energy $E$ as proportional to $\exp\left(-\frac{E}{k_BT}\right)$.\ \cite{gibbs} Then one can integrate to find a quantity proportional to the probability the energy exceeds $E$. For small $T$, this mirrors the simplest form of the Arrhenius equation where $E$ is the activation energy. For large $T$, however, the integrated form is more accurate as now the reaction rate diverges linearly with $T$.
\begin{equation}
\textnormal{reaction rate}=rk_BT\exp\left(-\frac{E}{k_BT}\right)=-rE+rk_BT+O\left(\frac{1}{T}\right)
\end{equation}
where $r$ is a constant with dimensions of inverse action.\footnote{According to the IUPAC, using a power of temperature in the pre-exponential factor is not novel.\ \cite{iupac} In their model however, $T$ is arbitrarily scaled and raised to any real power, probably resulting in overfitting of empirical data when they should really seek a more fundamental model.} This constant is dependent upon the concentration of the reactants (free nucleotides) so it cannot be known with any certainty for primordial kinetics. Nonetheless, it should be kept in mind that the purpose of this study is to determine the plausibility of phenomena so it will be sufficient to give a range of $r$ values for which the proposed reaction can progress. In the Monte Carlo simulation, the probability of a reaction occurring successfully will be $\textnormal{rate}\times\Delta t$ such that in the limit as $\Delta t\rightarrow0$ the average waiting time for an event is $1/\textnormal{rate}$ provided the rate remains constant (a Poisson process).

\subsection{Simulation results}

Figure 4 depicts the progression of this process using the ensemble of estimated constants that was described throughout Section 5.
\begin{figure}[b!]
\caption{RNA molecules vs. time on a log scale for a quasi-polymerase chain reaction modeled as Sections 4 and 5 describe it (blue). In purple is the same trial but removing the particles' mobility that is critical to quasi-PCR. To the right is the quasi-PCR efficacy after 100 seconds as a function of the rate constant that depends on physical factors around a prebiotic hydrothermal vent. Since this must be empirical, a range of plausible values is given as promised in subsection 5.4.}
\hspace{-.07in}\includegraphics[scale=.42]{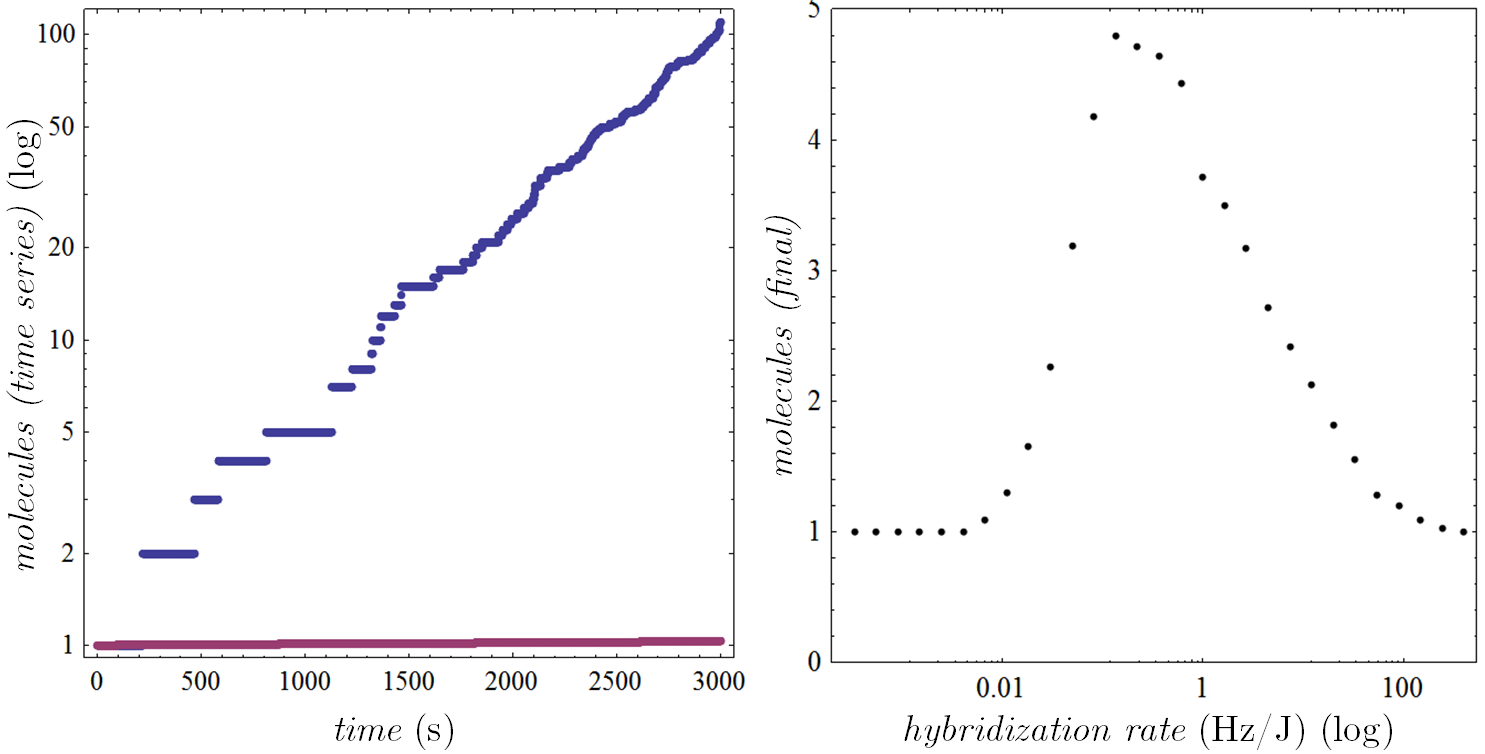}
\end{figure}
It is seen that at the constant vent temperature there is virtually no replication; after these 3000 seconds, the probability a molecule has been replicated is $\sim2\%$, averaged over 1000 runs of the process. Because by the end the true quasi-PCR simulation generates $\sim100$ independent polynucleotides that each become an effectively parallel computation, there was no need to examine many separate systems. The \textit{in silico} power of the proposed quasi-PCR establishes the reaction as a potential replication mechanism that deserves further examination.\footnote{The results section is brief due not to a lack of data collection and analysis but to an effort to emphasize the key finding: the possibility of progression of a quasi-polymerase chain reaction. Thorough analysis and discussion will be given in the next section.}

\section{Discussion: Quasi-PCR}

Mentioned in the introduction is ribozyme-discoverer Thomas Cech's view on the progression from complex organic molecules to an RNA world to an RNP world to the modern world as it is expressed in his 2011 paper.\ \cite{cech} He cites a ``likely self-replicating systems that preceded RNA'' and describes a potential replicating molecule that would be related to RNA. The molecular biology literature appears to be caught up in the specific chemistry that could have allowed repeated ligation of a polynucleotide. Here it is shown that the thermodynamics of the system could provide sufficient impetus for even reactions of high energy barrier to occur. As for pre-enzymatic reactivity, Baaske \textit{et al.}~\cite{baaske} describe how deep-sea thermal processes induce extreme accumulation of particles in vent pores, creating ideal settings for reactions. These successes show that the focus henceforth needs to be on the general physical factors that could have produced the inexplicable emergence of complexity. Indeed, the second plot in Figure 4 shows that a quasi-PCR can progress at a reasonable pace with the hybridization rate varying over four orders of magnitude. This suggests that the mechanism presented thus far can provide a more general framework for rapid polymerization in hydrothermal vent systems, irrespective of the detailed chemistry. On an even more fundamental level, this reaction is analogous to the classic Urey-Miller experiment \cite{miller} as they both involve systematically shocking small molecular components in hopes that several will exceed some high activation energy to ultimately yield a more ordered product.\footnote{Radiation-induced mutation provides yet another parallel of this process. Brief quantized excitations upset everything so that the system may land in a more favorable state. Why this is crucial when discussing the nature of life was discussed by Schr\"{o}dinger. \cite[pp.~42-45]{schrod}}

The mechanism presented here is particularly effective because it makes use of the large energy influx from a super-powerful 1300 horsepower hydrothermal vent. Compare the directness and minimal stochasticity of this hypothesis to that of Obermayer \textit{et al.} \cite{obermayer} which has the same purpose but depends on a subtle probabilistic tendency for hydrolytic cleavage to occur at unhybridized loci.\ \cite{usher} Though this did effectively induce a selective pressure for more complex RNA conformations allowing the investigators to see recurring sequence as well as structural motifs, there was no observation of strong exponential information replication akin to the one established in this paper.

\section{Applications: PCR Technology}

\subsection{Nonlinear PCR}

It should be noted \textit{a priori} that two arguments will be given and because one is the converse of the other, they must be thought of as independent assertions to avoid circularity.

A fundamental difference between true PCR as it is used in the laboratory and quasi-PCR is the nature of the thermal cycling: while the former is carried out in very discrete stages of denaturation, annealing, and elongation, the latter involves the periodic and continuous temperature oscillations shown in Figure 1.

\begin{wrapfigure}{l}{0.5\textwidth}
\begin{center}
\includegraphics[width=0.53\textwidth]{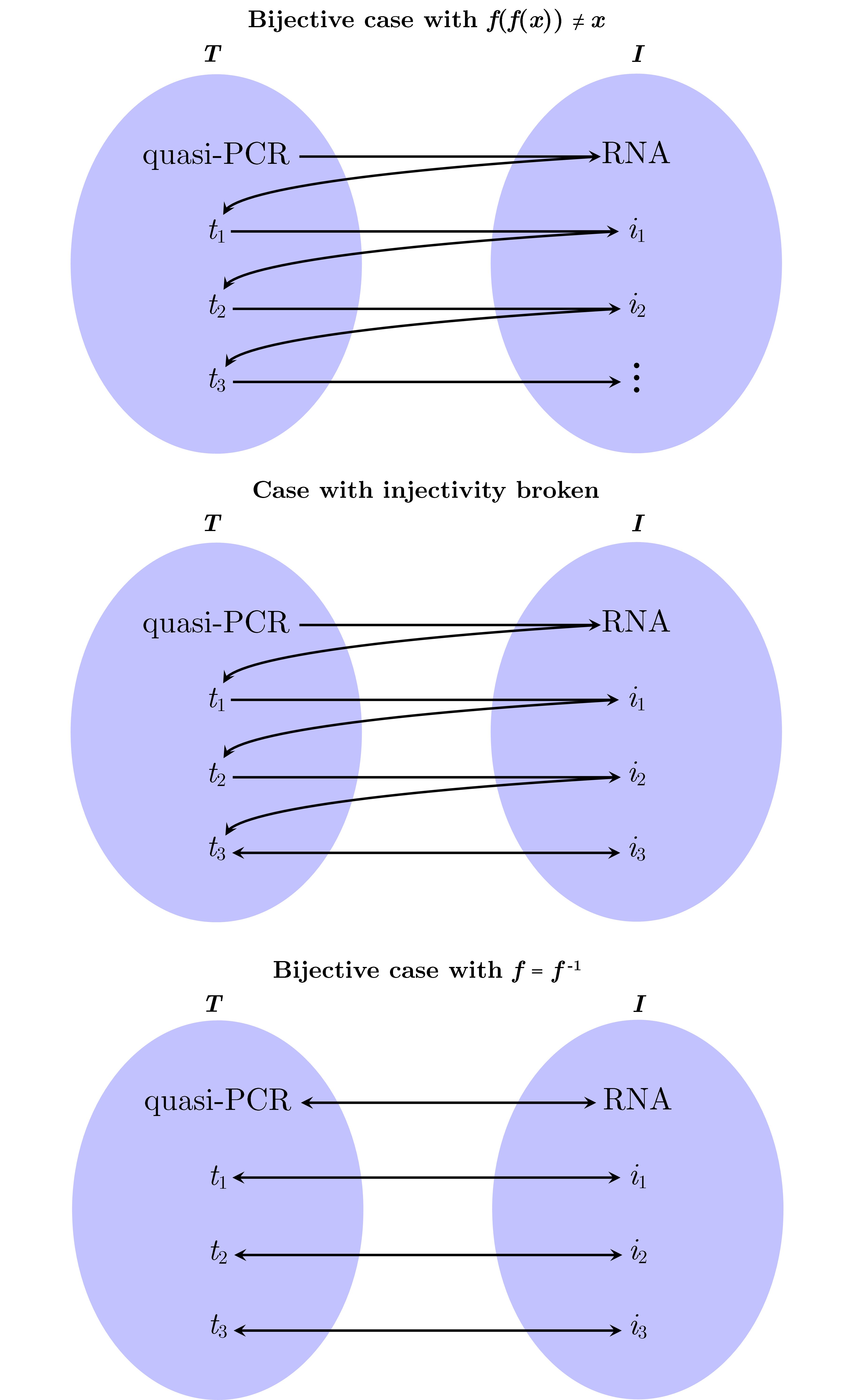}
\end{center}
\caption{The three kinds of possible maps formulated here between the spaces of thermodynamics and genetics. Biomimicry is possible if and only if the final bijective structure is the correct one.}
\end{wrapfigure}\vspace{-.2in}

The inevitable consequence of discrete cycling is that each stage must be maintained long enough so that the expected proportion of the molecules that are successful in whatever process is associated with that stage is greater than a certain near-unity threshold value. One familiar with mathematical optimization might find it absurd that the ideal temperature function out of all functions that map positive reals to positive reals would be a step function. Let ``nonlinear PCR'' be a PCR reaction whose temperature function was carefully extremized. The applicability of such a reaction will be supported by an argument of biomimicry of quasi-PCR.

\subsection{Conditions for biomimicry of quasi-PCR}

``Quasi-PCR'' and ``RNA'' will refer respectively to the thermodynamics of the quasi-PCR reaction as they are outlined in subsection 2.3 and to the extant genetic molecules. The underlying argument is as follows. Let $T$ denote the set of all possible thermal cycling phenomena and $I$ the set of all possible information transmitters. Thus we can say $\textnormal{quasi-PCR}\in T$ and $\textnormal{RNA}\in I$. Also let $f:T\rightarrow I$ and $I\rightarrow T$ meaning that given the argument of $f$ occurred, the output of $f$ is the member of the codomain that results in the most effective prebiotic replication. For the moment we take as a premise the quasi-PCR hypothesis: RNA came about in a PCR-like process with Figure 1's nonlinear temperature cycles. This would suggest that the hereditary molecules present today were those which most prevailed in quasi-PCR, ie. $f(\textnormal{quasi-PCR})=\textnormal{RNA}$. In order to justify biomimicry, the converse $f(\textnormal{RNA})$ must be determined. It will be shown that this equals quasi-PCR as one might expect only when bijectivity between $T$ and $I$ is assumed. Suppose $t_k$ and $i_k$ for $k=1,2,3,\dots$ are distinct members of $T$ and $I$ respectively excluding quasi-PCR and RNA. It was already argued that $\textnormal{quasi-PCR}\rightarrow\textnormal{RNA}$. $f(\textnormal{RNA})$ may take on one of two values: quasi-PCR (this is the bijective case), or something distinct represented by $t_1$. From here $t_1$ can map back to RNA (this is the $\textnormal RNA\leftrightarrow t_1$ case) or to a distinct $i_1$. Now $i_1$ cannot map back to quasi-PCR because that implies that a reaction between quasi-PCR and $i_1$ would be more effective than a reaction between $i_1$ and $t_1$ which must be more effective than $t_1$ with RNA (otherwise $t_1$ would map back to RNA but $i_1$ is distinct) which must be more effective than quasi-PCR with RNA (otherwise RNA would map back to quasi-PCR which was already the bijective case). Since $i_1$ cannot map to quasi-PCR, it must either map to $t_1$ (this is the $t_1\leftrightarrow i_1$ case) or to a distinct $t_2$. This method can be continued \textit{ad infinitum} and three types of structures will be left: bijectivities with inverse ($\textnormal{quasi-PCR}\leftrightarrow\textnormal{RNA}$ and $t_k\leftrightarrow i_k$ cases), chains of $t_1\rightarrow i_1\rightarrow t_2\rightarrow i_2\rightarrow\cdots$ interrupted by bijectivities, and chains that continue indefinitely. In practice, however, arbitrarily long chains cannot exist as at each link the replication must become more rapid but there must be a practical upper limit on its speed. With the two possible structures that remain, if bijectivity between the $T$ and $I$ can be shown, then $f$ must be its own inverse function establishing that $f(\textnormal{RNA})=\textnormal{quasi-PCR}$.

\subsection{Mapping $I$ to $T$}

The previous paragraph concluded that upon the premise of the quasi-PCR hypothesis, biomimicry of quasi-PCR is valid if and only if the following statement is accurate: \textit{If A is the thermal system that best replicates genetic molecule B, then B is the genetic molecule best replicated by A}. Here we develop a method to uniquely map information carriers to the thermodynamic system that would most encourage replication. There is no need to proceed \textit{de novo} as most of the requisite mathematics were already formulated in this paper in the analysis of quasi-PCR.

We wish to maximize the expected number of molecules that will be left after replication which is given by \eqref{mean}. This is equivalent to extremizing the functional $\int R(t)dt=\int\left(\frac{d}{dt}p_0^m+mr_-\right)dt$ which is a variational problem. $p_0$ is given in \eqref{p0}. The degradation rate $r_-$ will be the function subject to optimization since it can be explicitly expressed in terms of the temperature (see 5.4). The hybridization rate $r_+$ will be treated as a constant since we are now operating under the conditions of enzyme-catalyzed modern PCR. Since the inner function has no explicit time dependence, we can use Gelfand and Fomin's result that the Euler-Lagrange equation reduces to\ \cite[p.~19]{gelfand}
\begin{equation}\label{awful}
R-\frac{dr_-}{dt}\frac{\partial R}{\partial r_-'}=C\hspace{.08in}\textnormal{where}\hspace{.08in}R=\frac{d}{dt}\left[\frac{1-\frac{R}{r_-}\left(1-e^{-r_-/R}\right)}{1+r_-/r_+}\right]^m+mr_-\textnormal{ from }\eqref{p0}
\end{equation}
and $C$ is the constant obtained from reducing the order of the equation. The author is currently working to approximate an analytic solution to \eqref{awful} while preserving its critical features.\footnote{Necessary and sufficient simplification is apparently not an uncommon dilemma, as it is the subject of a preface by Bender \cite[pp.~2-4]{bender} and a whole book by Wolfram \cite[expressed concisely on p.~1025 in the context of biological evolution]{wolfram}.} Until then, preliminary Monte Carlo simulations for several temperature functions can be presented. Figure 6 depicts the average factor by which the number of molecules has been amplified for constant, square-wave, and quasi-PCR temperature functions. Note that simulations were only carried out for polynucleotides of length five and the results are expected to become even more dramatic for longer chains. The initial drop is due to the initial single molecule consistently undergoing denaturation among all the trials. It is the recovery from this depletion that occurs at differential rates among the paradigms.
\begin{figure}[b!]
\caption{Mean number of complete replications in simulated PCR with constant temperature $\sim 100^\circ C$ (blue), square wave temperature oscillations (purple), and quasi-PCR oscillations (gold). Averages are taken over 50 trials beginning from a single molecule. The latter method appears best, but analysis is by no means complete as there is no obvious way to put the paradigms on equal footing for a comparison when they are so fundamentally different.}
\hspace{-.07in}\includegraphics[scale=.57]{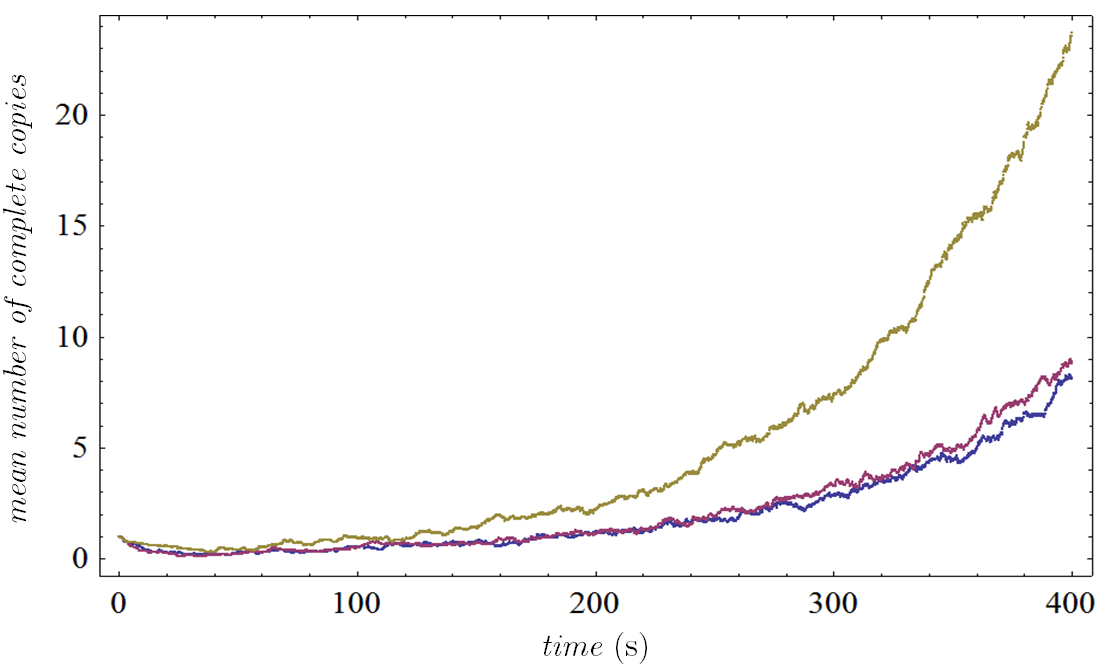}
\end{figure}

\subsection{Variational calculus over periodic functions}

It is a concern that the optimal temperature function according to \eqref{awful} may not be characteristic of a chain reaction---that is, it may not be periodic. A method is developed here to force periodicity when extremizing functionals that depend on the function and first time derivative.

Let the functional operating on temperature $T$ be $\mathcal{L}=\mathcal{L}\{T,T'\}$. For simplicity and since it is reasonable to start the chain reaction at the peak temperature, we will assume $T$ is an even function with angular frequency $\omega$. Using a Fourier cosine series we can rewrite the functional as
\begin{equation}
\mathcal{L}=\mathcal{L}\left\{\frac{a_0}{2}+\sum_{i=1}^{\infty}a_i\cos(i\omega t),-\omega\sum_{i=1}^{\infty}ia_i\sin(i\omega t)\right\}.
\end{equation}
Now this is an optimization of a countably infinite number of variables so we can resort to methods of traditional rather than variational calculus. In other words, all the derivatives of $\mathcal{L}$ with respect to the coefficients must identically vanish when the coefficients are optimal. Using the multivariable chain rule, when $j=1,2,3,\ldots$,
\begin{equation}
\frac{\partial\mathcal{L}}{\partial a_j}=\frac{\partial\mathcal{L}}{\partial T}\frac{\partial T}{\partial a_j}+\frac{\partial\mathcal{L}}{\partial T'}\frac{\partial T'}{\partial a_j}=\cos(j\omega t)\frac{\partial\mathcal{L}}{\partial T}-\omega j\sin(j\omega t)\frac{\partial\mathcal{L}}{\partial T'}=0.
\end{equation}
After rearrangement, the implicit function theorem allows the ratio of the partial derivatives to be reduced.
\begin{equation}
\frac{\partial\mathcal{L}/\partial T}{\partial\mathcal{L}/\partial T'}=-\left(\frac{\partial T'}{\partial T}\right)_{\!\mathcal{L}}=\omega j\tan(j\omega t)\hspace{.1in}\textnormal{for }j=1,2,3,\ldots
\end{equation}
where the subscript means $\mathcal{L}$ is being held constant as $T$ and its derivative are allowed to vary. Note the minus sign that is not a typo but is required by the implicit function theorem. Computation is likely easier using the less attractive $\partial_T\mathcal{L}/\partial_{T'}\mathcal{L}$. The optimization was done for the symbol temperature but of course it would generalize to $r_-$ when applied to \eqref{awful}.

\subsection{$R(t)$ for general chain reaction processes}

In a scenario more general than PCR, consider two processes $A$ and $B$ which occur in turn with respective temperature-dependent rates of $R_A(T)$ and $R_B(T)$. Let $s_0$ and $s_1$ be, respectively, the initial state and the state after process $A$ has occurred but before $B$. $B$ then is the transition from $s_1$ to $s_2$, making the reaction cyclic. Letting $\textbf{P}(t)$ be the probability vector over the system's states, the evolution has a simple master equation.
\begin{equation}\label{master}
\frac{d\textbf{P}}{dt}=\left(\begin{smallmatrix} -R_A & R_B \\
R_A & -R_B \end{smallmatrix}\right)\textbf{P}
\end{equation}
Given that all components begin in $s_0$ and that $R(t)=R_2(t)\times P[\textnormal{final state}]$ as described in subsection 4.2, the first-order linear system yields
\begin{equation}
R(t)=R_B(t)\int_0^tR_A(\tau)\exp\left(-\int_\tau^t(R_A(k)+R_B(k))dk\right)d\tau
\end{equation}
where the explicit temperature dependence of $R_A$ and $R_B$ are omitted. Again, the functional to be maximized is the integral of $R(t)$ from 0 to a constant so as to extremize \eqref{mean}.

\subsection{Applications support theory}

Here the independent converse argument mentioned in this section's introduction will be briefly described. The biomimicry proposal was premised upon the validity of the quasi-PCR hypothesis. Though evidence can be gathered for quasi-PCR, one can of course never be certain of what actually occurred in prebiotic times. Therefore one can never be certain whether nonlinear PCR is \textit{actually} mimicking anything that ever occurred in biology. What one \textit{can} plausibly be certain of is that our current biochemistry is such that it replicates very effectively when subjected to the thermal conditions of subsection 2.3. This then would serve as evidence that quasi-PCR was responsible for the emergence of lasting genetic memory. Figure 7 shows the relationship between these two arguments and highlights some of the peculiarities of the proposals made here. While the previous subsections made the unusual claim that hypothetical primordial biology applies to modern technology, this section is perhaps more obscure in its assertion that inferences about prebiotic systems can be made based on observations in biotechnology. This all stems from the unorthodox nature of the biomimicry presented here which is mimicking the physical environment in which the biological system is immersed, not the biology itself.
\begin{figure}
\caption{Given either the validity quasi-PCR or the favorability nonlinear PCR, the other is suggested. If the assumptions for both arguments are not carefully formulated they may appear circular.\vspace{.15in}}
\hspace{1.1in}\includegraphics[scale=.35]{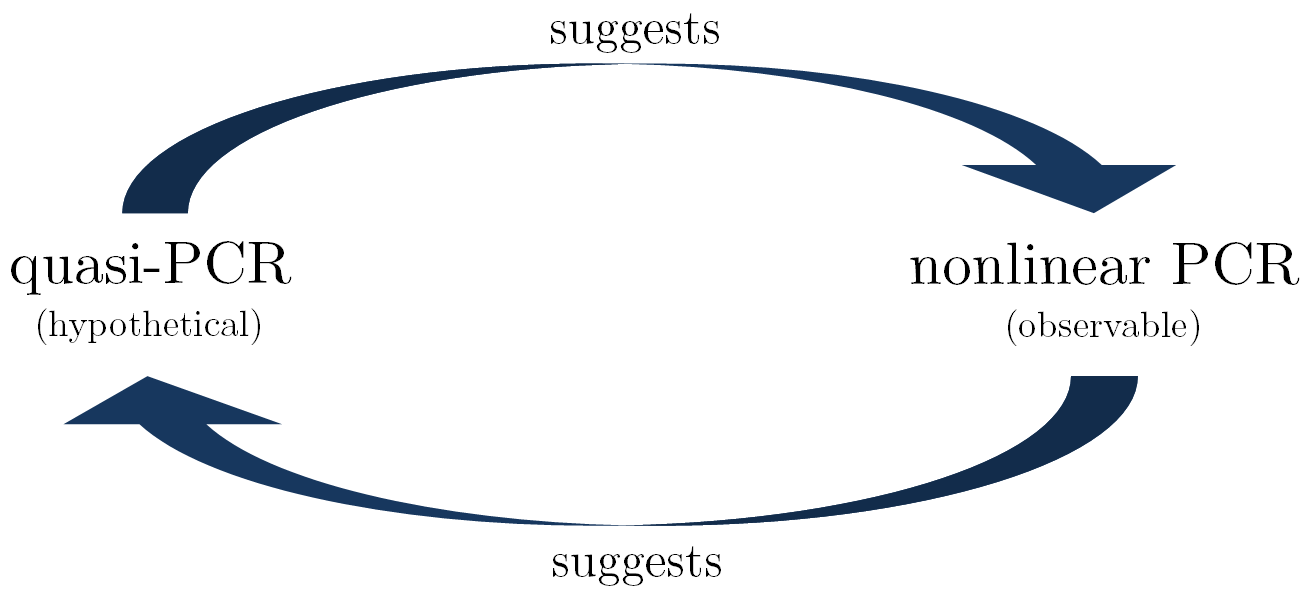}\end{figure}

\section{Conclusion}

A simulation of nucleotide flow, ligation, and cleavage encompassing all physical factors thought to be in play in prebiotic hydrothermal vents was conducted. Vent thermodynamics were shown to induce oscillations of temperature from each particle's perspective which result in a polymerase chain reaction-like process (quasi-PCR). Values for the rate constant for hydrogen bond degradation could not be obtained from the literature because a more precise denaturation model was derived in this paper that yields accurate kinetics for the large range of temperatures over which quasi-PCR operates. Nonetheless, it was shown that the reaction can progress effectively over several orders of magnitude of this constant. Thus one can be confident that reactions like the quasi-PCR \textit{could} proceed in prebiotic systems. However, due to the nature of studying the origin of life, one can never be certain if this was truly the reaction that provided the critical transition from inert organic compounds to the self-replicating information-storing pioneer molecules of the RNA world. This specificity is not needed for the results of this paper to be significant because, as discussed in section 6, quasi-PCR provides a very general framework for spontaneous polymerization that is fundamentally analogous to both the Urey-Miller experiment and diversity through radiation-induced mutation. The fact the quasi-PCR \textit{can} occur, regardless of whether it did motivates the conclusion that this instrumental transition in the origin of life without which the RNA world hypothesis has no credibility \textit{is} thermodynamically possible.

The biotechnology application was an unexpected implication of what began as a pursuit in pure biology. The initial computer simulations appear promising, but as discussed earlier, the ``experiments'' are were not yet carefully controlled. The most definitive conclusion on nonlinear PCR will be reached once the variational calculus approach (in progress) is completed.

\bibliographystyle{unsrt}

\begin{thebibliography}{99}

\bibitem{weaver}
Weaver, R. F., and Hedrick, P. W. \textit{Genetics}, Wm.\ C.\ Brown Publishers: Dubuque, IA 1995.
\bibitem{kruger}
Kruger, K., Grabowski, P.\ J., Zaug, A.\ J., Sands, J., Gottschling, D.\ E., and Cech, T.\ R. 1982. Self-splicing RNA: autoexcision and autocyclization of the ribosomal RNA intervening sequence of \textit{Tetrahymena}. \textit{Cell} \textbf{31}, 147-157.
\bibitem{crick}
Crick, F.\ H.\ C. 1968. The origin of the genetic code. \textit{J. Mol. Biol.} \textbf{38}, 367-379.
\bibitem{orgel}
Orgel, L.\ E. 1968. Evolution of the genetic apparatus. \textit{J. Mol. Biol.} \textbf{38}, 381-393.
\bibitem{gilbert}
Gilbert, W. 1986. Origin of life: the RNA world. \textit{Nature} \textbf{319}, 618.
\bibitem{cech}
Cech, T.\ R. 2011. The RNA worlds in context. \textit{Cold Spring Harb. Perspect. Biol.} \textbf{4}, 7.
\bibitem{joyceorgel}
Joyce, G.\ F.\ and Orgel, L.\ E. \textit{Prospects for understanding the origin of the RNA world}. The RNA World, Cold Spring Harbor Laboratory Press: New York, NY 1993, p. 13.
\bibitem{baaske}
Baaske, P., Weinert, F.\ M., Duhr, S., Lemke, K.\ H., Russell, M.\ J., and Braun, D. 2007. Extreme accumulation of nucleotides in simulated hydrothermal pore systems. \textit{Proc. Natl. Acad. Sci. USA} \textbf{104}, 9346-9351.
\bibitem{chen7}
Chen, I.\ A., Roberts, R.\ W., and Szostak, J.\ W. 2007. The emergence of competition between model protocells. \textit{Science} \textbf{305}, 1474-1476.
\bibitem{obermayer}
Obermayer, B., Krammer, H., Braun, D., and Gerland, U. 2011. Emergence of information transmission in a prebiotic RNA reactor. \textit{Phys. Rev. Lett.} \textbf{107}, 018101.
\bibitem{chen12}
Chen, I.\ A. and Nowak, M.\ A. 2012. From prelife to life: how chemical kinetics become evolutionary dynamics. \textit{Acc. Chem. Res.}
\bibitem{krammer}
Krammer, H., M\"{o}ller, F.\ M., Braun, D. 2012. Thermal, autonomous replicator made from transfer RNA. \textit{Phys. Rev. Lett.} \textbf{108}, 238104.
\bibitem{pagani}
Pagani, M., Lemarchand, D., Spivack, A., Gaillardet, J. 2005. A critical evaluation of the boron isotope-\textit{p}H proxy: the accuracy of ancient ocean \textit{p}H estimates. \textit{Geochim. Cosmochim. Ac.} \textbf{69}, 953-961.
\bibitem{pinti}
Pinti, D.\ L. \textit{The Origin and Evolution of the Oceans}. M. Gargaud \textit{et al.} (Ed.) Lectures in Astrobiology, Vol. I, Springer-Verlag: Berlin 2005 (pp.~83-112).
\bibitem{powner}
Powner, M. W., Gerland, B., Sutherland, J. D. 2009. Synthesis of activated pyrimidine ribonucleotides in prebiotically plausible conditions. \textit{Nature} \textbf{459}, 239-242.
\bibitem{robertson}
Robertson, M.\ P. and Joyce, G.\ F. 2012. The origins of the RNA world. \textit{Cold Spring Harb. Perspect. Biol.} \textbf{4}, 5.
\bibitem{shubin}
Shubin, M. A. \textit{Foundations of the Classical Theory of Partial Differential Equations}, Volume 1, Viniti: Moscow 1988.
\bibitem{fetter}
Fetter, A. L. and Valecka, J. D. \textit{Theoretical Mechanics of Particles and Continua}, Courier Dover Publicatons: Chemsford, MA 2003.
\bibitem{duhr}
Duhr, S., Braun, D. 2006. Why molecules move along a temperature gradient. \textit{Proc. Natl. Acad. Sci. USA} \textbf{103} No. 52, pp. 19678-19682.
\bibitem{cussler}
Cussler, E.\ L. \textit{Diffusion: Mass Transfer in Fluid Systems}, Cambridge University Press: Cambridge 1997.
\bibitem{hamming}
Hamming, R. W. \textit{Numerical Methods for Scientists and Engineers}, Dover Publications: New York, NY 1973.
\bibitem{haase}
Haase, K. M., \textit{et al.} 2007. Young volcanism and related hydrothermal activity at 5$^\circ$S on the slow-spreading southern Mid-Atlantic Ridge. \textit{Geochem. Geophys. Geosyst.} \textbf{8}, Q11002, doi:10.1029/2006GC001509.
\bibitem{santos}
Berberan-Santos, M.\ N. Green's function method and the first-order linear differential equation. \textit{J. Math. Chem.} \textbf{48}, 175-178.
\bibitem{turin}
Kobayashi, H., Mark, B.\ L., and Turin, W. \textit{Probability, Random Processes, and Statistical Analysis}, Cambridge University Press: Cambridge 2011.
\bibitem{bansal}
Bansal, R. \textit{Fundamentals of engineering electromagnetics}, CRC Press: Boca Raton, FL 2006.
\bibitem{little}
Little, S.\ A., Stolzenbach, K.\ D., and Von Herzen, R.\ P. 1987. Measurements of plume flow from a hydrothermal vent field. \textit{J. Geophys. Res.} \textbf{92}, 2587-2596.
\bibitem{sengers}
Sengers, J.\ V. and Watson, J.\ T.\ R. 1986. Improved international formulations for the viscosity and thermal conductivity of water substance. \textit{J. Phys. Chem. Ref. Data} \textbf{15}, 1291-1314.
\bibitem{kos}
Koschinsky, A., Garbe-Sch\"{o}nberg, D., Sander, S., Schmidt, K., Gennerich, H., and Strauss, H. 2008. Hydrothermal venting at pressure-temperature conditions above the critical point of seawater, 5$^\circ$S on the Mid-Atlantic Ridge. \textit{Geol.} \textbf{36}, 615-618.
\bibitem{zhao}
Zhao, Y.\ H., Abraham, M.\ H., and Zissimos, A.\ M. 2003. Fast calculation of van der Waals volume as a sum of atomic and bond contributions and its application to drug compounds. \textit{J. Org. Chem.} \textbf{68}, 7368-7373.
\bibitem{einstein}
Einstein, A. 1905. \"{U}ber die von der molekularkinetischen theorie der w\"{a}rme geforderte bewegung von in ruhenden fl\"{u}ssigkeiten suspendierten teilchen. \textit{Ann. Phys. (Berlin)} \textbf{322}, 549-560.
\bibitem{mit}
Southard, J. Lecture, Special Topics: An introduction to fluid motions, sediment transport, and current-generated sedimentary structures, Chapter 3. Massachusetts Institute of Technology: Cambridge, MA 2006.
\bibitem{gibbs}
Gibbs, J.\ W. \textit{Elementary Principles in Statistical Mechanics}, Scribner: New York, NY 1902.
\bibitem{iupac}
IUPAC. \textit{Compendium of Chemical Terminology}, 2nd ed. (the "Gold Book"), Blackwell Scientific Publications: Oxford 1997.
\bibitem{miller}
Miller, S.\ L. 1953. A production of amino acids under possible primitive Earth conditions. \textit{Science} \textbf{117}, 528-529.
\bibitem{usher}
Usher, D. and McHale, A. 1976. Hydrolytic stability of helical RNA: a selective advantage for the natural 3',5'-bond. \textit{Proc. Natl. Acad. Sci. USA} \textbf{73}, 1149-1153.
\bibitem{schrod}
Schr\"{o}dinger, E. \textit{What is Life?}, Cambridge University Press: Cambridge 1992.
\bibitem{gelfand}
Gelfand, I.\ M.\ and Fomin, S.\ V. \textit{Calculus of Variations}, Dover Publications: New York, NY 1963.
\bibitem{bender}
Bender, A.\ E. \textit{An Introduction to Mathematical Modeling}, Wiley: New York, NY 1978.
\bibitem{wolfram}
Wolfram, S. \textit{A New Kind of Science}, Wolfram Media: Champaign, IL 2002.

\end{thebibliography}

\end{document}